\documentclass[12pt]{article} 
\usepackage{graphicx,subfigure,latexsym,amssymb}
\usepackage{bm}
\usepackage{amsmath}
\usepackage{mathtools}

\newcommand{\be}{\begin{equation}}
\newcommand{\ee}{\end{equation}}
\newcommand{\bear}{\begin{eqnarray}}
\newcommand{\eear}{\end{eqnarray}}
\newcommand{\ba}{\begin{array}}
\newcommand{\ea}{\end{array}}

\def\({\left(}
\def\){\right)}

\begin{document}

\begin{titlepage}
\vfill
\begin{flushright}
\end{flushright}

\vfill
\begin{center}
{\Large\bf Relaxation Times for Chiral Transport Phenomena and Spin Polarization in Strongly Coupled Plasma }

\vskip 0.7in
Shiyong Li\footnote{e-mail: {\tt  sli72@uic.edu}} and
Ho-Ung Yee\footnote{e-mail:
{\tt hyee@uic.edu}}
\vskip 0.3in

 {\it Department of Physics, University of Illinois,} \\
{\it Chicago, Illinois 60607 }\\[0.15in]

\end{center}

\vfill

\begin{abstract}

We compute the dynamical relaxation times for chiral transport phenomena in strongly coupled regime using the AdS/CFT correspondence. These relaxation times can be a useful proxy for the dynamical time scale for achieving equilibrium spin-polarization of quasi-particles in the presence of magnetic field and fluid vorticity. We identify the Kubo relations for these relaxation times and clarify some previous issues regarding time-dependence of the Chiral Vortical Effect. We study the consequences of imposing time-reversal invariance on parity-odd thermal noise fluctuations that are related to chiral transport coefficients by the fluctuation-dissipation relation. We find that time-reversal invariance dictates the equality between some of the chiral transport coefficients as well as their relaxation times.


\end{abstract}

\vfill

\end{titlepage}
\setcounter{footnote}{0}

\baselineskip 18pt \pagebreak
\renewcommand{\thepage}{\arabic{page}}
\pagebreak

\section{Introduction \label{sec1}}

    The anomalous transport phenomena arising from chiral anomaly that could be present in the chiral-symmetry restored phase of quark-gluon plasma created in ultra-relativistic heavy-ion collisions have attracted much attention recently (see the reviews \cite{Kharzeev:2013ffa,Kharzeev:2015znc,Skokov:2016yrj,Kharzeev:2012ph,Miransky:2015ava,Hattori:2016emy} and the references therein). One such phenomenon, the Chiral Magnetic Effect (CME) \cite{Fukushima:2008xe,Vilenkin:1980fu} predicts a dipole charge separation \cite{Kharzeev:2007tn,Kharzeev:2007jp,Hirono:2014oda,Jiang:2016wve} which is proportional to the magnetic field produced in off-central collisions along the direction perpendicular to the reaction plane \cite{Skokov:2009qp,Bzdak:2011yy,Deng:2012pc,Bloczynski:2012en,Tuchin:2014iua,McLerran:2013hla}, that could result in the observable charge dependence of angular correlations of two charged pions \cite{Voloshin:2004vk,Abelev:2009ac,Abelev:2009ad,Adamczyk:2015eqo,Adam:2015vje}. A closely related phenomenon, the Chiral Vortical Effect(CVE) \cite{Erdmenger:2008rm,Banerjee:2008th} may also induce additional separation of charged hadrons proportional to the vorticity or the kinetic angular momentum of the plasma fireball \cite{Kharzeev:2010gr,Jiang:2015cva}. 
Both of these effects rely on the fluctuating chiral charge (or axial charge) imbalance between the right-handed and the left-handed quarks that could arise either from initial state fluctuations of color charge density \cite{Kharzeev:2001ev} or from statistical sphaleron transitions in the quark-gluon plasma \cite{Mace:2016svc,Mace:2016shq}. These effects also induce interesting gapless propagating modes of chiral charges, the Chiral Magnetic Wave \cite{Kharzeev:2010gd,Newman:2005hd}, which may induce a charge quadrupole moment proportional to the magnetic field \cite{Burnier:2011bf,Gorbar:2011ya,Yee:2013cya}.
To clearly disentangle the proposed signatures of these effects from the other sources that constitute the background \cite{Asakawa:2010bu,Bzdak:2009fc,Wang:2009kd,Pratt:2010zn,Bzdak:2013yla}, the iso-bar heavy-ion collision experiments comparing Rb and Zr that have 10\% difference in the strength of magnetic field are currently on-going in the Relativistic Heavy-Ion Collider at the Brookhaven National laboratory \cite{Skokov:2016yrj,Deng:2016knn}.
The evidence for the CME has been observed recently in the condensed matter systems of Dirac\cite{Li:2014bha} and Weyl\cite{Huang:2015eia,Zhang:2016ufu} semi-metals where quasi-electrons feature similar concepts of chirality found in relativistic massless fermions.

In the limit of static and homogeneous magnetic field and vorticity, the magnitude of the CME and the CVE conductivities defined by 
\be
J^\mu_{chiral}=\sigma_B B^\mu+\sigma_V \omega^\mu\,,\quad B^\mu\equiv {1\over 2}\epsilon^{\mu\nu\alpha\beta}u_\nu F_{\alpha\beta}\,,\quad \omega^\mu \equiv \epsilon^{\mu\nu\alpha\beta}u_\nu\nabla_\alpha u_\beta\,,
\ee
is completely fixed by chiral anomaly. For the simplest case of a theory with one right-handed chiral fermion species, we have\footnote{For QCD with $N_F$ light flavors of Dirac quarks, the net current is a sum of left- and right-handed currents both of which should be multiplied by $N_c N_F$ and the chemical potentials are the chiral ones, $\mu_{L/R}=\mu_V\mp \mu_A$. The effects for the left-handed currents carry the additional opposite sign compared to those for the right-handed currents.} 
\be 
\sigma_B = \frac{\mu}{4\pi^2}\,,\quad \sigma_V={\mu^2\over 8\pi^2}\,,
\ee
where $\mu$ is the (chiral) chemical potential. For $\sigma_V$, there exists additional correction of order $T^2$ where $T$ is temperature, that is related to mixed chiral-gravitational anomaly \cite{Landsteiner:2011cp,Landsteiner:2011iq}. These results are topologically protected and not modified by interactions, which has been confirmed in many different approaches, for instance, in weakly coupled perturbation theory \cite{Kharzeev:2009pj,Jimenez-Alba:2015bia,Satow:2014lva}, the AdS/CFT correspondence \cite{Yee:2009vw,Rebhan:2009vc,Gynther:2010ed,Hoyos:2011us,Bu:2016oba,Ammon:2017ded}, hydrodynamics with the second law of thermodynamics \cite{Son:2009tf,Kharzeev:2011ds,Loganayagam:2011mu}, effective field theories \cite{Jensen:2012kj,Jensen:2013rga,Sadofyev:2010is}, and the chiral kinetic theory \cite{Son:2012wh,Stephanov:2012ki, Chen:2012ca,Son:2012zy,Chen:2014cla,Manuel:2014dza,Chen:2015gta,Mueller:2017arw,Hidaka:2016yjf,Huang:2018wdl}. 

In out-of equilibrium conditions, these conductivities depend more on microscopic interactions and generally deviate from their equilibrium values. The question of particular interests is how these conductivities behave when the magnetic field and the vorticity is time-dependent with a frequency $\omega$:
\be
J^\mu_{chiral}(\omega)=\sigma_B(\omega)B^\mu(\omega)+\sigma_V(\omega)\omega^\mu(\omega)\,,\label{freq}
\ee
The computation of the frequency dependence of the CME has been done both in weakly coupled\cite{Kharzeev:2009pj,Son:2012zy,Kharzeev:2016sut} and strongly coupled regimes \cite{Yee:2009vw,Amado:2011zx,Landsteiner:2013aba}, and the results are not universal anymore.
In the quasi-particle picture of these effects at weak coupling \cite{Kharzeev:2016sut}, 
the results at finite frequency depend on the relaxation dynamics of spin polarization of chiral fermions along the direction of magnetic field or vorticity, in addition to the more topological effect from the Berry phase that is responsible for the equilibrium values \cite{Son:2012wh,Stephanov:2012ki, Chen:2012ca}. For example, there appears a new contribution to the current from spin magnetization at finite frequency \cite{Kharzeev:2016sut}. 

Another way of looking at this is that the frequency-momentum dependence of chiral conductivities is related to the parity (P)-odd thermal noise fluctuations via the fluctuation-dissipation theorem \cite{Mamo:2015xkw}. We will review this connection in more detail in section \ref{T-reversal} where we derive the general consequences on these chiral conductivities when we impose the time-reversal (T) invariance of the theory on the P-odd thermal noise fluctuations\footnote{Note that a right-handed Weyl fermion remains right-handed under time-reversal transformation since helicity which is a product of momentum and spin is T-even. Without a CP-violating complex phase, a chiral (gauge) theory is invariant under time reversal and CP.}. In the simplest case of current-current noise correlation, one derives the fluctuation-dissipation relation
\be
\langle \delta \bm{J}^i(k) \delta \bm{J}^j(-k)\rangle\equiv G^{ij}_{rr}(k)=\left({1\over 2}+n_B(\omega)\right)\rho^{ij}(k)\,,\label{FD}
\ee
in terms of the spectral density $\rho^{ij}(k)\equiv G_{ra}^{ij}(k)-G_{ar}^{ij}(k)=G_{ra}^{ij}(k)+(G_{ra}^{ji}(k))^*$ where the last equality is a consequence of hermiticity of current operators (that is equivalent to the real-valuedness of the retarded correlation function in coordinate space) and the translational invariance of the system. Note that $\rho^{ij}(k)$ is a hermitian matrix in $(i,j)$ but needs not be real-valued. This hermiticity of $\rho^{ij}(k)$ together with translational invariance is sufficient to guarantee that the noise correlation is real-valued in coordinate space as it should be. Recalling that the retarded correlation function is $G_R^{ij}(k)=-i G_{ra}^{ij}(k)$, the P-odd part of the retarded current correlation function
\be
G_R^{ij}(k)=i\tilde\sigma_B(k)\epsilon^{ijl}{\bm k}^l\,,
\ee
that is responsible for frequency-momentum dependent chiral magnetic effect\footnote{As will be seen in section \ref{Kubo}, this $\tilde\sigma_B(k)$, although related to, is not precisely equal to the conductivity $\sigma_B(\omega)$ appearing in the constitutive relation (\ref{freq}) due to the energy-momentum shear-diffusion pole.},
gives a P-odd contribution to the spectral density \cite{Mamo:2015xkw}
\be
\rho^{ij}(k)\sim -2 \,{\rm Im}[\tilde\sigma_B(k)] i\epsilon^{ijl}{\bm k}^l\equiv \rho^{\rm odd}(k)i\epsilon^{ijl}{\bm k}^l\,,
\ee
in terms of a real-valued function $\rho^{\rm odd}(k)=-2\,{\rm Im}[\tilde\sigma_B(k)]$.
The reality of retarded correlation function in coordinate space gives a constraint $\tilde\sigma_B(k)=(\tilde\sigma_B(-k))^*$, which dictates that
\be
\rho^{\rm odd}(-k)=-\rho^{\rm odd}(k)\,,
\ee
especially the P-odd spectral density vanishes in zero momentum limit. 
When ${\bm k}\to 0$, we have $\rho^{\rm odd}(\omega)=2\xi_5\omega+\cdots$ in small $\omega$ limit, where the transport coefficient $\xi_5$ roughly correspond to a second order correction to the CME \cite{Kharzeev:2011ds,Megias:2014mba}\footnote{This is not precise. The more precise such ``Kubo relation'' will be derived in section \ref{Kubo}.}
\be
J^\mu_{CME}=\sigma_B B^\mu+\xi_5 (u\cdot \nabla) B^\mu\,,
\ee
 and together with $n_B(\omega)\to {T\over \omega}$ in (\ref{FD}) in $\omega\ll T$, we have the P-odd thermal noise fluctuations in hydrodynamics regime,
\be
\langle \delta \bm{J}^i(\bm{x},t)\delta \bm{J}^j(\bm{x'},t')\rangle=2T\xi_5 \epsilon^{ijk}{\partial\over\partial \bm{x}^k}\delta(\bm{x}-\bm{x'})\delta(t-t')\,.
\ee 
Since the thermal noise spectrum in general should depend on real-time microscopic dynamics of thermal ensemble, they can't be topological and should be non-universal. This is consistent with the fact that the above P-odd spectral density is insensitive to the topologically protected value of $\lim_{\bm {k}\to 0}\lim_{\omega\to 0}{\rm Re}[\tilde\sigma_B(k)]={\mu\over 4\pi^2}$ and only depends on the non-universal part of ${\rm Im}[\tilde\sigma_B(k)]$. The $\xi_5$ was computed in two-flavor perturbative QCD in leading log as \cite{Jimenez-Alba:2015bia}
\be
\xi_5\approx -{0.5\over \alpha_s^2\log(1/\alpha_s)}{\sigma_B\over T}\,.
\ee

These frequency-dependent chiral conductivities are also important in any realistic simulation of heavy-ion collisions in an attempt to quantify the effects from these anomalous transport phenomena on the experimental observables \cite{Hirono:2014oda,Jiang:2016wve}.

A useful characterization of frequency-dependent transport coefficients is the relaxation time defined by
\be
\sigma(\omega)=\sigma(0)(1+i\omega \tau)+{\cal O}(\omega^2)\,,
\ee
which is motivated from a simple relaxation time pole expanded in first order in $\omega$,
\be
\sigma(\omega)=\sigma(0){i/\tau\over\omega+i/\tau}\,,
\ee
or equivalently, from a Israel-Stewart type treatment expanded in first order in time derivative $u\cdot\nabla$,
\be
(u\cdot\nabla)J^\mu_{CME}=-{1\over\tau_B}(J^\mu_{CME}-\sigma_B B^\mu)\longrightarrow\quad J^\mu_{CME}=\sigma_B B^\mu-\sigma_B \tau_B (u\cdot\nabla)B^\mu+\cdots\,.
\ee
We should note however that the real response of the system may not be simply given by the Israel-Stewart form \cite{Denicol:2011fa}, and in that case the relaxation time defined as above should instead be viewed as a characteristic dynamical time scale of the problem that appears in second order hydrodynamics \cite{Baier:2007ix}. 

The full anomalous transport phenomena in the anomaly frame \cite{Loganayagam:2011mu,Loganayagam:2012pz,Landsteiner:2012kd} or no-drag frame \cite{Stephanov:2015roa,Rajagopal:2015roa} consist of four transport coefficients in leading order of derivatives
\bear
T^{\mu\nu}_{chiral}&=&\sigma_\epsilon^B(B^\mu u^\nu+B^\nu u^\mu)+\sigma_\epsilon^V(\omega^\mu u^\nu+\omega^\nu u^\mu)\,,\nonumber\\
J^\mu_{chiral}&=& \sigma_B B^\mu +\sigma_V\omega^\mu\,.\label{leading}
\eear
The two additional transport coefficients, $\sigma_\epsilon^{B/V}$, are responsible for the anomalous energy-flow or momentum density along the magnetic field and vorticity. Their values are also fixed by chiral anomaly to be (up to temperature corrections) \cite{Loganayagam:2012pz,Landsteiner:2012kd}
\be
\sigma_\epsilon^B={\mu^2\over 8\pi^2}\,,\quad \sigma_\epsilon^V={\mu^3\over 6\pi^2}\,.
\ee
We correspondingly introduce the four relaxation times in the next order in time derivative\footnote{Using the equation of motion of ideal hydrodynamics and the Bianchi identity for field strength tensor, one can replace the time derivative $u\cdot\nabla$ with spatial derivatives. But, this is equivalent up to equations of motion, and does not affect the Kubo relations derived from the equation of motion.}
\bear
T^{\mu\nu}_{chiral}&=&\sigma_\epsilon^B(B^\mu-\tau_\epsilon^B(u\cdot\nabla)B^\mu)u^\nu+\sigma_\epsilon^V(\omega^\mu-\tau_\epsilon^V(u\cdot\nabla)\omega^\mu) u^\nu+(\mu\leftrightarrow \nu)\,,\nonumber\\
J^\mu_{chiral}&=& \sigma_B( B^\mu-\tau_B(u\cdot\nabla)B^\mu) +\sigma_V(\omega^\mu-\tau_V(u\cdot\nabla)\omega^\mu)\,.\label{nextleading}
\eear
In this work, we compute these relaxation times, $(\tau_\epsilon^{B/V},\tau_{B/V})$, in strongly coupled regime using the AdS/CFT correspondence.

The dynamics of spin polarization that underlies the above relaxation times is more general beyond the massless chiral limit and is applicable even for massive spinful particles such as Lambda baryon. In theoretical understanding of recently measured $\Lambda$ hyperon spin polarization in off-central heavy-ion collisions at RHIC \cite{STAR:2017ckg}, it is vitally important to know the dynamical time scale of this spin polarization, that is, the relaxation time it takes to achieve the spins polarized in a finite fluid vorticity or magnetic field. Therefore, by studying the relaxation times of the anomalous chiral transport phenomena in massless chiral limit, we can get a useful proxy for this important time scale at the quark level before the hadronization happens where the quark spin polarization is transferred to the $\Lambda$ hyperon spin polarization. 
This is another motivation we have in this study.

In section \ref{T-reversal} we show that the time-reversal (T) invariance of a microscopic theory imposed on the P-odd thermal noise fluctuations of momenta and currents requires
\be
\sigma_\epsilon^B=\sigma_V\,,\quad \tau_\epsilon^B=\tau_V\,.
\ee
The first equality has been observed before in several different approaches \cite{Loganayagam:2012pz,Landsteiner:2012kd}, but its connection to the T-invariance seems novel. The second equality is our new finding.
We confirm this in our numerical computation in the AdS/CFT correspondence within 
the numerical accuracy of our analysis.

The frequency dependence of the CVE was previously considered in Refs.\cite{Amado:2011zx,Landsteiner:2013aba} in both weakly and strongly coupled regimes. 
Writing the P-odd part of the retarded current-momentum correlation function as
\be
\langle \bm{J}^i(k) T^{0j}(-k)\rangle_R =i\tilde\sigma_V(k)\epsilon^{ijl}\bm{k}^l\,,
\ee
it was observed that $\tilde\sigma_V(k)$ in the hydrodynamic regime behaves as
\be
\tilde\sigma_V(k)=\left(\sigma_V-{n\over\epsilon+p}\sigma_\epsilon^V {\omega\over \omega+i\gamma_\eta \bm{k}^2}\right){i\gamma_\eta \bm{k}^2\over \omega+i\gamma_\eta \bm{k}^2}\,,\label{tildeV}
\ee
where $n$ is the charge density and $\gamma_\eta=\eta/(\epsilon+p)$ is the shear-diffusion constant. The physics behind the appearance of the shear-diffusion pole structure was also explained in Ref.\cite{Landsteiner:2013aba}: it is a consequence of the energy-momentum conservation Ward identity.
From this expression, it was observed that the retarded correlation function $\tilde\sigma_V(k)$ vanishes in $\bm{k}\to 0$ limit for all $\omega\neq 0$, while the strict $\omega=0$ limit reproduces the correct CVE coefficient $\sigma_V$. 

There is nothing wrong in these observations, but we point out that this retarded correlation function, $\tilde\sigma_V(k)$, is {\it not directly equal} to the frequency-momentum dependent {\it transport coefficient}, $\sigma_V(k)$, appearing in the constitutive relations: $\tilde\sigma_V(k)\neq\sigma_V(k)$. One way of seeing why these observations are not related to $\sigma_V(\omega)$ is that (\ref{tildeV}) is derived from the constitutive relation at leading order, that is (\ref{leading}), without any relaxation time or higher-order corrections in the constitutive relations. To extract the relaxation time appearing in the transport coefficient $\sigma_V(\omega)$ from the retarded correlation function $\tilde\sigma_V(k)$, we need to work with the constitutive relations at next leading order (\ref{nextleading}) including the relaxation times to derive the correct Kubo relations between the relaxation time and the correlation function $\tilde\sigma_V(k)$. We perform this exercise in section \ref{Kubo}.

Our paper is organized as follows. In section \ref{Kubo} we derive the correct Kubo relations between the relaxation times of chiral transport phenomena and the retarded correlation functions of energy-momenta and currents. In section \ref{T-reversal} we study the consequences of imposing the time-reversal invariance on the P-odd thermal noise fluctuations that can be obtained from the retarded correlation functions via the fluctuation-dissipation relation: we find that T-invariance dictates $\tau_\epsilon^B=\tau_V$. We then proceed in section \ref{sec3} to the numerical computation of these relaxation times in the AdS/CFT correspondence.
We conclude with discussions in section \ref{sec4}.

\section{Kubo relations for the relaxation times of chiral transport phenomena \label{Kubo}}

Hydrodynamics is an effective theory describing dynamics of a thermal system on large length and time scales. 
The relevant degrees of freedom are the local velocity $u^\mu(x)$ (satisfying $u\cdot u=-1$) and the local densities of conserved quantities such the energy $\epsilon(x)$ and a charge $n(x)$. The hydrodynamic equations of motion are given by the local conservation laws $ \nabla_\mu T^{\mu \nu}=0$, $\nabla_\mu J^{\mu}=0 $, together with constitutive relations which express $T^{\mu \nu}$ and $J^{\mu}$ in terms of the local hydrodynamics variables themselves. These constitutive relations are organized by the number of spatial derivatives involved. The lowest order is the ideal hydrodynamics, and the viscous corrections start to appear in the first order of derivatives. 
In the case of a chiral fluid, new terms of chiral transport phenomena proportional to fluid vorticity and external magnetic field must be present in the constitutive relations in order to ensure the second law of the thermodynamics\cite{Son:2009tf,Loganayagam:2011mu}
\bear
T^{\mu\nu}&=&(\epsilon+p)u^\mu u^\nu+p g^{\mu\nu}+\tau^{\mu\nu}+\tau^{\mu\nu}_{chiral}\,,\nonumber\\
J^\mu&=&n u^\mu + \nu^\mu+\nu^\mu_{chiral}\,,
\eear
where our metric is mostly positive, $p=p(\epsilon,n)$ is the pressure, and $n$ is the charge density. The viscous and chiral transport terms are
\bear
\tau^{\mu\nu}&=&- \eta \Delta^{\mu\alpha}\Delta^{\nu\beta} (\nabla_\alpha u_\beta + \nabla_\beta u_\alpha- \frac{2}{3}g_{\alpha\beta}\nabla_\lambda u^\lambda )- \zeta \Delta^{\mu\nu} \nabla_\lambda u^\lambda\,,\nonumber\\
\tau^{\mu\nu}_{chiral}&=&\sigma_\epsilon^B(B^\mu-\tau_\epsilon^B(u\cdot\nabla)B^\mu)u^\nu+\sigma_\epsilon^V(\omega^\mu-\tau_\epsilon^V(u\cdot\nabla)\omega^\mu) u^\nu+(\mu\leftrightarrow \nu)\,,\nonumber\\
\nu^\mu&=&-\sigma T\Delta^\mu (\mu/T)+\sigma E^\mu\,,\nonumber\\
\nu^\mu_{chiral}&=& \sigma_B( B^\mu-\tau_B(u\cdot\nabla)B^\mu) +\sigma_V(\omega^\mu-\tau_V(u\cdot\nabla)\omega^\mu)
\eear
where we include only the relaxation time terms of our interest at the next leading order in derivatives, since the other second order corrections do not interfere with the above relaxation time terms in our final Kubo relations. Our notations are conventional, and $\Delta^\mu=(u^\mu u^\nu+g^{\mu\nu})\nabla_\nu\equiv \Delta^{\mu\nu}\nabla_\nu$ is the spatially projected derivative in local rest frame,
$E^\mu = F^{\mu\nu}u_\nu$, $B^\mu= \frac{1}{2}\epsilon^{\mu\nu\alpha\beta}u_\nu F_{\alpha \beta}$ are the electromagnetic fields, and $ \omega^\mu = \epsilon^{\mu\nu\alpha\beta}u_\nu\partial_\alpha u_\beta $ is the fluid vorticity. 
The $\sigma_B\tau_B$ is equivalent to one of the second order transport coefficients $-\xi_5$ studied in Refs.\cite{Kharzeev:2011ds,Jimenez-Alba:2015bia}.
The electric conductivity and charge diffusion terms proportional to the conductivity $\sigma$ will not affect our Kubo relations for the chiral transport coefficients. In the presence of external background fields, the hydrodynamic equations take the following form,
\be
\nabla_\mu T^{\mu\nu} = F^{\nu\alpha}J_\alpha\,,\quad
\nabla_\mu J^\mu = C E_\mu B^\mu \,, \quad C={1\over 4\pi^2}\,.
\ee

In the ``anomaly'' \cite{Loganayagam:2011mu,Loganayagam:2012pz,Landsteiner:2012kd} or ``no-drag'' \cite{Stephanov:2015roa} frame, the leading order chiral transport coefficients are given by
\be
\sigma_B=C\mu\,,\quad \sigma_V=\sigma_\epsilon^B={1\over 2}C\mu^2\,,\quad\sigma_\epsilon^V={2\over 3}C\mu^3\,,
\ee
up to temperature corrections which are related to the mixed chiral-gravitational anomaly \cite{Landsteiner:2011cp,Landsteiner:2011iq,Jensen:2012kj}. One can also choose to work in the Landau frame \cite{Son:2009tf} where one removes the chiral transports in the energy-momentum tensor by redefining the fluid velocity as
\be
u^\mu=u^\mu_{LF}-{1\over\epsilon+p}\left(\sigma_\epsilon^B(B^\mu-\tau_\epsilon^B(u\cdot\nabla)B^\mu)+\sigma_\epsilon^V(\omega^\mu-\tau_\epsilon^V(u\cdot\nabla)\omega^\mu) \right)\,.\label{shift}
\ee
Note that this upsets the naive counting of derivative expansions.
This will then shifts the transport coefficients $\sigma_{B/V}$ and $\tau_{B/V}$ appearing in $\nu^\mu_{chiral}$ in the Landau frame as
\be
\sigma_{B/V}^{LF}=\sigma_{B/V}-{n\over\epsilon+p}\sigma_\epsilon^{B/V}\,,\quad \tau_{B/V}^{LF}=\tau_{B/V}-{n\over\epsilon+p}\tau_\epsilon^{B/V}\,.\label{relation}
\ee

Recall that these two ``frames'' (or any other frame) are different {\it descriptions} of the {\it same system} : a change of variables. Any final physics results, such as the retarded correlation functions of operators or the spectrum of thermal fluctuations, should be the same independent of the choice of frames, {\it provided} we used the correct relations (\ref{relation}) between the transport coefficients appearing in different frames.
Note that the fluid velocity is not such a physics result: it is a descriptive variable. While we will work in the anomaly/no-drag frame in the following, the same retarded correlation functions and the Kubo relations are reproduced in the Landau frame as well. In checking this explicitly, it is necessary to keep the induced second-order term in the Landau frame coming from the shear viscosity term when we make the shift in fluid velocity (\ref{shift});
\be
-{\eta\over\epsilon+p}\Delta^{\mu\alpha}\Delta^{\nu\beta}\left(\sigma_\epsilon^B\nabla_\alpha B_\beta+\sigma_\epsilon^V \nabla_\alpha\omega_\beta+(\alpha\leftrightarrow\beta)-{\rm trace}\right)\,.
\ee
The necessity of this higher order term in showing the equivalence between the two frames was observed in Ref.\cite{Yee:2014dxa}, and the basic reason is a violation of derivative counting in (\ref{shift}).

To obtain the Kubo relations for these chiral transport coefficients \cite{Amado:2011zx,Landsteiner:2012kd}, we consider small fluctuations of metric and gauge field, $(h_{\mu\nu},A_\mu)$ in the background of static equilibrium plasma in flat spacetime. Furthermore, we restrict ourselves to the effects that are linear in these fluctuations $h_{\mu\nu}$ and $A_\mu$. These fluctuations will drive the fluid away from equilibrium, which is described by the fluid velocity $u^\mu = (1, u^i(t,\bf x))$ (valid to linear order in $u^i$) as well as $\delta \epsilon$ and $\delta n$, that are determined by solving the hydrodynamics equation of motion. For our purpose, it is enough to switch on just $(h_{tx}(t, y),h_{tz}(t, y))$ and $(A_x(t, y), A_z(t, y))$ which depend only on $(t,y)$. In the frequency-momentum Fourier space of $e^{-i\omega t + iky}$, the constitutive relations that are linear in the fluctuations become 
\bear 
T^{tt} &=& \epsilon+\delta \epsilon\,,\nonumber\\
T^{ti} &=& (\epsilon+p)u^i + p h_{ti} - \sigma^B_\epsilon (1 + i \omega \tau^B_\epsilon) {\epsilon}^{ijk}\partial_j A_k - \sigma^V_\epsilon (1 + i \omega \tau^V_\epsilon){\epsilon}^{ijk}\partial_j(u^k+h_{tk})\,,\nonumber\\
T^{ij} &=& (p+c_s^2 \delta\epsilon)\delta^{ij}-\eta(\partial_i u^j+\partial_j u^i-\frac{2}{3}\delta_{ij}\partial_k u^k)\,,\nonumber\\
J^i &=& n u^i -\sigma_B(1+ i \omega \tau_B) {\epsilon}^{ijk}\partial_j A_k -\sigma_V(1 + i \omega \tau_V){\epsilon}^{ijk}\partial_j (u^k + h_{tk})\,,
\eear
where $c_s$ is the speed of sound. For the sake of simplicity, we drop the bulk viscosity $\zeta$.
The relevant hydrodynamic equations of motion from the energy-momentum conservation become
\bear
&&\partial_t\delta \epsilon+(\epsilon_0+P_0)\partial_i u^i = 0\,,\nonumber\\
&&(\epsilon + p)\partial_t u^i+ c_s ^2\partial_i \delta\epsilon-\eta(\partial^2 u^i + \partial_j\partial_i u^j - \frac{2}{3}\partial_i\partial_k u^k) -\sigma^B_\epsilon(1+i \omega \tau^B_\epsilon) {\epsilon}^{ijk} \partial_t\partial_j A_k \nonumber\\&&
-\sigma^V_\epsilon(1+i \omega \tau^V_\epsilon) {\epsilon}^{ijk} \partial_t\partial_j(u^k+ h_{tk})  =-(\epsilon+ p)\partial_t h_{ti} -n \partial_t A_i\,,
\eear
from which we can solve for the velocity, the energy density and the charge density fluctuations,
\bear
\delta\epsilon&=& 0 \,,\quad \delta n=0\,,\nonumber\\
u^x&=&\frac{ - h_{tz}\sigma^V_\epsilon(1+i \omega \tau^V_\epsilon) i \omega k  \frac{i \gamma_\eta \bm{k}^2}{\epsilon + p} + h_{tx} \omega (\omega + i \gamma_\eta \bm{k}^2) + A_x \omega \frac{n }{\epsilon 
 + p} (\omega + i \gamma_\eta \bm{k}^2)}{-(\omega+ i \gamma_\eta \bm{k}^2)^2} \nonumber\\
&+& \frac{A_z \frac{i \omega k}{\epsilon + p}( \sigma^B_\epsilon(1 + i \omega \sigma^B_\epsilon)(\omega + i \gamma_\eta \bm{k}^2) - \frac{n}{\epsilon + p} \omega\sigma^V_\epsilon (1 + i \omega \tau^V_\epsilon))}{(\omega+ i \gamma_\eta \bm{k}^2)^2}\,,\nonumber\\
u^z&=&\frac{ + h_{tx}\sigma^V_\epsilon(1+i \omega \tau^V_\epsilon) i \omega k  \frac{i \gamma_\eta \bm{k}^2}{\epsilon + p} + h_{tz} \omega (\omega + i \gamma_\eta \bm{k}^2) + A_z \omega \frac{n }{\epsilon 
 + p} (\omega + i \gamma_\eta \bm{k}^2)}{-(\omega+ i \gamma_\eta \bm{k}^2)^2} \nonumber\\
&-& \frac{A_x \frac{i \omega k}{\epsilon + p}( \sigma^B_\epsilon(1 + i \omega \sigma^B_\epsilon)(\omega + i \gamma_\eta \bm{k}^2) - \frac{n}{\epsilon + p} \omega\sigma^V_\epsilon (1 + i \omega \tau^V_\epsilon))}{(\omega+ i \gamma_\eta \bm{k}^2)^2}\,.
\eear
Inserting these back into the constitutive relations, we obtain the induced energy-momentum tensor and the currents that are linear in the fluctuating background $(h_{\mu\nu}, A_\mu)$.
The retarded two point functions of the energy-momenta and currents can then be obtained by differentiating these with respect to the metric and gauge field fluctuations $(h_{\mu\nu}, A_\mu)$. 

This computation for the P-odd parts of the energy-momenta and currents retarded correlation functions 
is a small extension of what has already been done in Ref.\cite{Landsteiner:2013aba}, now including the relaxation time terms at next leading order.
Writing 
\bear
\langle \bm {J}^i(k)\bm{J}^j(-k)\rangle_R&=&i\tilde\sigma_B(k)\epsilon^{ijl}{\bm k}^l\,,\quad \langle \bm{J}^i(k) T^{0j}(-k)\rangle_R=i\tilde\sigma_V(k)\epsilon^{ijl}{\bm k}^l\,,\nonumber\\
\langle T^{0i}(k)\bm{J}^j(-k)\rangle_R&=&i\tilde\sigma^B_\epsilon(k)\epsilon^{ijl}{\bm k}^l\,,\quad \langle T^{0i}(k) T^{0j}(-k)\rangle_R=i\tilde\sigma^V_\epsilon(k)\epsilon^{ijl}{\bm k}^l\,,\label{retcor2}
\eear
we finally obtain the retarded correlation functions including the relaxation times
\bear
\tilde\sigma_B(k)&=& \sigma_B (1 + i \omega \tau_B)- \frac{n}{\epsilon +p}\bigg(\sigma^B_\epsilon(1+ i \omega \tau^B_\epsilon)+\sigma_V(1+ i \omega \tau_V) \nonumber\\
&-&{\sigma^V_\epsilon}(1 + i \omega \tau^V_\epsilon) \frac{n}{\epsilon + p}\frac{\omega}{\omega + i \gamma_\eta \bm{k}^2}\bigg)\frac{\omega}{\omega + i \gamma_\eta \bm{k}^2} \,,\nonumber\\ 
\tilde\sigma_V(k)&=& \left(\sigma_V (1 + i \omega \tau_V) - \frac{n}{\epsilon + p} \sigma^V_\epsilon (1 + i \omega \tau^V_\epsilon) \frac{\omega}{\omega + i \gamma_\eta \bm{k}^2} \right)\frac{i \gamma_\eta \bm{k}^2}{\omega + i \gamma_\eta \bm{k}^2}\,,\nonumber\\ 
\tilde\sigma^B_\epsilon(k)&=& \left(\sigma^B_\epsilon (1 + i \omega \tau^B_\epsilon) - \frac{n}{\epsilon + p} {\sigma^V_\epsilon} (1 + i \omega \tau^V_\epsilon) \frac{\omega}{\omega + i \gamma_\eta \bm{k}^2} \right) \frac{i \gamma_\eta \bm{k}^2}{\omega + i \gamma_\eta \bm{k}^2}\,,\nonumber\\
\tilde\sigma_\epsilon^V(k)&=&\sigma^V_\epsilon (1+ i \omega \tau^V_\epsilon) \frac{-\gamma_\eta^2 \bm{k}^4}{(\omega + i \gamma_\eta \bm{k}^2)^2}\,,\label{fullret}
\eear
where $\gamma_\eta={\eta/(\epsilon+p)}$ is the shear diffusion constant. The appearance of shear-diffusion poles in the above retarded correlation functions was explained in Ref.\cite{Landsteiner:2013aba} and is due to the energy-momentum Ward identity.
In the zero frequency limit, all four chiral transport coefficients decouple from each other, and these give the correct Kubo relations for the leading order chiral transport coefficients, $(\sigma_{B/V},\sigma_\epsilon^{B/V})$. 
On the other hand, in the other limit of $\bm k\to 0$ with $\omega\neq 0$, the $\langle\bm{J}T\rangle_R$, $\langle T\bm{J}\rangle_R$ and $\langle TT \rangle_R$ correlators all vanish due to the diffusion pole structure. To extract the relaxation times $(\tau_{B/V},\tau_\epsilon^{B/V})$ from these retarded correlation functions, it is clear that these diffusion  pole structures must be stripped off first before taking appropriate $\bm{k}\to 0$ limit and looking at the remaining linear term in $\omega$.  For example, the Kubo relation for the relaxation time $\tau_\epsilon^V$ can be
\be
\tau_\epsilon^V=-{1\over\sigma_\epsilon^V}\lim_{\omega\to 0}{1\over\omega} \lim_{\bm k\to 0}{\rm Im}\left[ \frac{(\omega + i \gamma_\eta \bm{k}^2)^2}{-\gamma_\eta^2 \bm{k}^4}\tilde\sigma_\epsilon^V(k)\right]\,,
\ee
and we can write down the Kubo relations for other relaxation times:
\bear
\tau_V&=&{1\over\sigma_V}\lim_{\omega\to 0}{1\over\omega} \lim_{\bm k\to 0}{\rm Im}\left[ \frac{\omega + i \gamma_\eta \bm{k}^2}{i\gamma_\eta \bm{k}^2}\tilde\sigma_V(k)+\frac{n}{\epsilon + p} \sigma^V_\epsilon (1 + i \omega \tau^V_\epsilon) \frac{\omega}{\omega + i \gamma_\eta \bm{k}^2}\right]\,,
\eear
and similar expressions for $\tau_\epsilon^{B}$ and $\tau_B$.
Practically, we compute $\tau_\epsilon^V$ first and use it in the Kubo relations for the $(\tau_V,\tau_\epsilon^B)$ to compute them, and finally compute $\tau_B$. For these Kubo relations, we need to know $\gamma_\eta=\eta/(\epsilon+p)$ precisely, and 
fortunately in the AdS/CFT correspondence we can easily achieve this due to the Kovtun-Son-Starinets result on the shear viscosity \cite{Kovtun:2004de}, $\eta=s/4\pi$, that allows us to compute $\gamma_\eta$ precisely.
Our numerical study in the AdS/CFT correspondence that will be described in more detail in section \ref{sec3} shows that these Kubo relations can be successfully used to extract the relaxation times of chiral transport phenomena.

\section {   Consequences of time-reversal (T) invariance     \label{T-reversal}}

In this section, we would like to derive some interesting consequences of having time-reversal (T) invariance in the microscopic theory. The T-invariance is quite generic in typical chiral (gauge) theories, since a T- (or equivalently CP-) violation could only arise from a non-removable complex phase in the Lagrangian, which is not easy to have in general.
Recall that the chirality (or helicity) that comes from a spin-momentum alignment remains invariant under T-transformation. The ensemble with a finite chiral chemical potential does not violate T explicitly, since chiral charge and chemical potential are T-even quantities. The chiral anomaly $\partial_\mu J^\mu=C E\cdot B$ is also consistent with T-symmetry as the both sides are T-odd. We would like to impose the T-invariance of the microscopic theory on the  random noise fluctuations around equilibrium {\it a la} Onsager, whose strength is related to the spectral density via fluctuation-dissipation relation. In turn, the spectral density is obtained from transport coefficients in hydrodynamics regime, and this gives T-invariance constraints on the transport coefficients. We will follow this line of steps for the P-odd thermal noise fluctuations, the P-odd spectral densities, and the chiral transport coefficients of our interests.
The importance of T-invariance in chiral transport phenomena was previously emphasized in Ref.\cite{Kharzeev:2011ds}.

Let us first review the general consequences of T-invariance on the random noise fluctuations of hermitian operators $\hat{\cal O}^a$ :
\be
\langle \delta {\cal O}^a (t_1)\delta {\cal O}^b(t_2)\rangle\equiv \langle {\cal O}^a_r(t_1) {\cal O}^b_r(t_2)\rangle={1\over Z} {\rm Tr}\left(e^{-\beta( \hat H-\mu\hat N)}{1\over 2}\{\hat{\cal O}^a(t_1),\hat{\cal O}^b(t_2)\}\right)\,.\label{flucdef}
\ee
The hatted objects are the quantum operators, and the objects without hat are the field variables in the Schwinger-Keldysh path integral. We use the ``ra''-notation where $r={1\over 2}(1+2)$ and $a=(1-2)$ with $(1,2)$ meaning the two contours of forward and backward times. Also, $Z={\rm Tr}(e^{-\beta(\hat H-\mu\hat N)})$, the anti-commutator is $\{\hat A,\hat B\}=\hat A\hat B+\hat B\hat A$, and $\hat {\cal O}(t)=e^{i\hat H t}\hat {\cal O}e^{-i\hat H t}$.
Using the hermiticity of $\hat{\cal O}^a$, it is easy to see that the fluctuation correlation functions are real-valued
\be
\langle {\cal O}^a_r(t_1) {\cal O}^b_r(t_2)\rangle^*=\langle {\cal O}^a_r(t_1) {\cal O}^b_r(t_2)\rangle\,.\label{real}
\ee
The T-invariance of the theory means that the hamiltonian is invariant under complex conjugation up to a unitary transformation\footnote{The notion of complex conjugation of an operator depends on the basis of states. This ambiguity is taken care of by additional unitary transformation $\hat U$.}
\be
\hat H^*=\hat U^\dagger\hat H\hat U\,.\label{Hinv}
\ee
Each physical operator is also assigned a well-defined T-parity,
\be
\hat U (\hat{\cal O}^a)^* \hat U^\dagger =\tau_a \hat {\cal O}^a\,,\quad\tau_a=\pm 1\,.
\ee
For example, it is intuitively clear that the momentum density $\hat T^{0i}$ and the charge currents $\hat J^i$ should have T-parity $\tau=-1$, while the charge density $\hat N$ has T-parity $\tau=+1$.
From the above two equations it is easy to derive
\be
\hat U (\hat{\cal O}^a(t))^* \hat U^\dagger =\tau_a \hat {\cal O}^a(-t)\,.\label{negt}
\ee
By taking complex conjugation of (\ref{flucdef}) and using the fact that it is real (\ref{real}), we have
\bear
\langle {\cal O}^a_r(t_1) {\cal O}^b_r(t_2)\rangle&=&\langle {\cal O}^a_r(t_1) {\cal O}^b_r(t_2)\rangle^*={1\over Z} {\rm Tr}\left(e^{-\beta( \hat H^*-\mu\hat N^*)}{1\over 2}\{(\hat{\cal O}^a(t_1))^*,(\hat{\cal O}^b(t_2))^*\}\right)\nonumber\\
&=&\tau_a\tau_b{1\over Z} {\rm Tr}\left(e^{-\beta( \hat H-\mu\hat N)}{1\over 2}\{\hat{\cal O}^a(-t_1),\hat{\cal O}^b(-t_2)\}\right)=\tau_a\tau_b\langle {\cal O}^a_r(-t_1) {\cal O}^b_r(-t_2)\rangle\,,\nonumber\\
\eear
where we use (\ref{Hinv}) and (\ref{negt}) in going from the first line to the second.
In terms of the frequency space correlation function defined by
\be
G_{rr}^{ab}(\omega)=\int_{-\infty}^\infty dt\,e^{i\omega t}\,\langle {\cal O}^a_r(t) {\cal O}^b_r(0)\rangle\,,
\ee
the T-invariance requires that
\be
G_{rr}^{ab}(-\omega)=\tau_a\tau_b G_{rr}^{ab}(\omega)\,.\label{Tcon}
\ee

We next consider the fluctuation-dissipation relation that can be proved in general by Lehmann-type representation
\be
G^{ab}_{rr}(\omega)=\left({1\over 2}+n_B(\omega)\right)\left(G^{ab}_{ra}(\omega)-G^{ab}_{ar}(\omega)\right)\,,\label{FD2}
\ee
where the retarded correlation function is related to the ``ra''-correlation function by $G^{ab}_R=-i G_{ra}^{ab}$.
Up to this point, the indices $(a,b)$ could label the space position $\bm x$ of the operators, 
but let us make the space position of the operators more explicit in the following, and $(a,b)$ refer only to the operator species. We then introduce the Fourier momentum space by
\be
G_{ra}^{ab}(\omega,\bm k)\equiv G_{ra}^{ab}(k)=\int {d^3\bm x} \,e^{-i\bm k\cdot\bm x}\,G_{ra}^{ab}(\omega,\bm x)\,,
\ee
and the above relations (\ref{Tcon}) and (\ref{FD2}) hold true in the Fourier momentum space as well. Note our notation $k=(\omega,\bm k)$. 
The retarded correlations function in time and position coordinates can easily be shown to be real-valued as they should:
\be
G^{ab}_R(t,\bm x)=-i\theta(t){1\over Z}{\rm Tr}\left(e^{\beta(\hat H-\mu\hat N)} [\hat {\cal O}^a(t,\bm x),\hat{\cal O}^b(0)]\right)\,.
\ee
This implies in the Fourier space, $G^{ab}_R(-k)=(G^{ab}_R(k))^*$, which is equivalent to 
\be
G^{ab}_{ra}(-k)=-(G^{ab}_{ra}(k))^*\,.\label{realcond}
\ee
From the definition, we have 
\be
G^{ab}_{ra}(t,\bm x)=\langle {\cal O}^a_r(t,\bm x) {\cal O}^b_a(0)\rangle=\langle {\cal O}^a_r(0) {\cal O}^b_a(-t,-\bm x)\rangle=\langle {\cal O}^b_a(-t,-\bm x) {\cal O}^a_r(0)\rangle=G^{ba}_{ar}(-t,-\bm x)\,,
\ee
where we use the translation invariance in space-time in the second equality.
In the Fourier space, this translates to 
\be
G^{ab}_{ar}(k)=G^{ba}_{ra}(-k)=-(G^{ba}_{ra}(k))^*\,.
\ee
where the last equality comes from (\ref{realcond}). Using this result, the fluctuation-dissipation relation (\ref{FD2}) becomes
\be
G^{ab}_{rr}(k)=\left({1\over 2}+n_B(\omega)\right)\left(G^{ab}_{ra}(k)+(G^{ba}_{ra}(k))^*\right)\equiv \left({1\over 2}+n_B(\omega)\right)\rho^{ab}(k)\,,\label{spectdensity}
\ee
with the spectral density $\rho^{ab}(k)$ that is a hermitian matrix in terms of the indices $(a,b)$. Note that the spectral density can have an imaginary part when it is anti-symmetric in $(a,b)$. Indeed, this is what happens for the P-odd spectral functions arising from the P-odd retarded correlation functions of chiral transport phenomena \cite{Mamo:2015xkw}. This is consistent with the fact that the fluctuation correlation functions in space-time coordinates $G^{ab}_{rr}(t,\bm x)$ is real-valued.
In terms of the retarded correlation functions, the spectral density is given by
\be
\rho^{ab}(k)=i\left(G^{ab}_R(k)-(G^{ba}_R(k))^*\right)\,.\label{retspec}
\ee

The T-invariance constraint (\ref{Tcon}) and the relation (\ref{spectdensity}) gives the constraint on the spectral density.
Since
\be
\left({1\over 2}+n_B(-\omega)\right)= -\left({1\over 2}+n_B(\omega)\right)\,,
\ee
we have the T-invariance constraint
\be
\rho^{ab}(-\omega,\bm k)=-\tau_a\tau_b \rho^{ab}(\omega,\bm k)\,,
\ee
which in turn gives the constraints on the retarded correlation functions via (\ref{retspec}).

We are now ready to apply this to the P-odd retarded correlation functions (\ref{retcor2}) that we obtained in the hydrodynamics regime in the previous section. The interesting case is when ${\cal O}^a={\bm J}^i$ and ${\cal O}^b=
T^{0j}$, both of which have T-parity $\tau=1$. From (\ref{retcor2}) we have
\be
G_R^{ab}(k)=i\tilde\sigma_V(k)\epsilon^{ijl}{\bm k}^l\,,\quad G^{ba}_R(k)=-i\tilde\sigma_\epsilon^B(k)\epsilon^{ijl}{\bm k}^l\,,
\ee
which gives
\be
\rho^{ab}(k)=\left(-\tilde\sigma_V(k)+(\tilde\sigma_\epsilon^B(k))^*\right)\epsilon^{ijl}{\bm k}^l\,,
\ee
and we have the T-invariance constraint
\be
\tilde\sigma_V(\omega,\bm{k})-(\tilde\sigma_\epsilon^B(\omega,\bm {k}))^*=-\tilde\sigma_V(-\omega,\bm{k})+(\tilde\sigma_\epsilon^B(-\omega,\bm {k}))^*\,.\label{Tconeq}
\ee
Using the expression (\ref{fullret}) of $\tilde\sigma_V(k)$ and $\tilde\sigma_\epsilon^B(k)$ in the hydrodynamics regime, 
a short algebra results in the constraint
\be 
(\sigma_V(1 + i \omega \tau_V) - \sigma_\epsilon^B(1 + i \omega \tau^B_\epsilon))\frac{1}{\omega + i \gamma_\eta k^2} = (\sigma_V(1 - i \omega \tau_V) - \sigma_\epsilon^B(1 - i \omega \tau^B_\epsilon))\frac{1}{\omega - i \gamma_\eta k^2}\,,
\ee
which should be satisfied
for arbitrary frequency and momentum. This is true {\it if and only if} the following T-invariance constraints are met
\be
\sigma_V = \sigma_\epsilon^B\,,\quad \tau_V=\tau_\epsilon^B\,.
\ee
Our numerical computation in the AdS/CFT correspondence in the next section confirms these relations.

We can obtain a stronger version of T-invariance constraint using rotational invariance which tells that 
\be
\tilde\sigma_V(k)=\tilde\sigma_V(\omega,|{\bm k}|)\,,\quad \tilde\sigma_\epsilon^B(k)=\tilde\sigma_\epsilon^B(\omega,|{\bm k}|)\,.\label{rotinvnc}
\ee
The reality of retarded correlation functions in coordinate space dictates 
\be
\tilde\sigma_V(-k)=\left(\tilde\sigma_V(k)\right)^*\,,\quad \tilde\sigma_\epsilon^B(-k)=\left(\tilde\sigma_\epsilon^B(k)\right)^*\,,
\ee
and together with (\ref{rotinvnc}), the T-invariance constraint (\ref{Tconeq}) becomes
\be
{\rm Re}\left[\tilde\sigma_V(\omega,|\bm k|)\right]={\rm Re}\left[\tilde\sigma_\epsilon^B(\omega,|\bm k|)\right]\,.
\ee
The Kramers-Kronig dispersion relation of retarded correlation functions due to causality then implies that the imaginary parts of $\tilde\sigma_V(k)$ and $\tilde\sigma_\epsilon^B(k)$ are the same as well, and hence we finally arrive at the T-invariance constraint 
\be
\tilde\sigma_V(k)=\tilde\sigma_\epsilon^B(k)\,,
\ee
which is valid even beyond the hydrodynamics regime.

We close this section with a short digression to Weyl semi-metals. In these materials the origin of pseudo-chiral massless dispersion relation
for left-handed and right-handed quasi particles are separated in the Bloch momentum space. 
Since the chirality is T-even while the momentum flips sign under T, this separation in momentum space breaks T-invariance. It would be therefore consistent to have $\sigma_V\neq\sigma_\epsilon^B$ in these condensed matter systems.
It seems that this is indeed the case \cite{Landsteiner:2013sja}\footnote{We thank Karl Landsteiner for pointing this out to us.}.

\section{Relaxation times of chiral transport phenomena in the AdS/CFT correspondence \label{sec3}}


In this section we take a minimal version of holographic model in the AdS/CFT correspondence, where the chiral anomaly manifests as a 5-dimensional Chern-Simons term of the holographic gauge field that corresponds to the $U(1)_R$ chiral symmetry of a right-handed chiral fermion.
The 5-dimensional action is 
\be
S=\frac{1}{16 \pi G} \int d^5x \sqrt{-g}\left(R+12-\frac{1}{4}F_{AB}F^{AB}+\frac{\kappa}{3}\epsilon^{MNPQR}A_M F_{NP}F_{QR}\right)\,,
\ee
where the Chern-Simons coefficient $\kappa$ should be $\kappa=G/(2\pi)$ to match the chiral anomaly for a single right-handed Weyl fermion. The $F_{MN}=\nabla_MA_N-\nabla_N A_M$ is the field strength of the $U(1)$ gauge field $A_M$ (capital letters run for 5-dimensional coordinates, and the greek letters for 4-dimensional coordinates). This theory is T-invariant\footnote{Under T-transformation, the spatial components of gauge field are odd while the time component is even, as can be inferred from their coupling to the charge current.}.

The corresponding equations of motion are,
\bear
G_{AB} - 6 g_{AB}+\frac{1}{2}F_{AC}F^C_B + \frac{1}{8} g_{AB} F^2 &=& 0\,,\nonumber\\
\nabla_B F^{BA}+ \kappa \epsilon^{AMNCD}F_{MN}F_{CD} &=& 0\,,
\eear
where $g_{AB}$ is the metric, $G_{AB} = R_{AB}-\frac{1}{2}g_{AB}R$ is the five dimensional Einstein tensor. The equations of motion allow an exact charged black-brane solution, which describes a finite temperature plasma with non-zero chemical potential $\mu$, 
\be 
ds^2 = r^2(-f(r)dt^2 + d\vec x^2) + \frac{1}{r^2 f(r)} dr^2 \,, \quad    A_t= -\frac{\mu r_H^2}{r^2}\,,
\ee
where $f(r)=1-\frac{m}{r^4}+\frac{q^2}{r^6}$ is the blackening factor, and $r_H$ is the horizon where $f(r_H)=0$. The parameter $q$ is related to the chemical potential by $\sqrt{3}q={\mu r_H^2}$, and the Hawking temperature is $T={r_H^2 f'(r_H)\over 4\pi}$. The equation of state of this fluid is
\be 
p(T,\mu)=\frac{m(T,\mu)}{16\pi G}\,,
\ee
and all other thermodynamic quantities are derived from the pressure. For our purpose, we need
\be
\gamma_\eta={\eta\over\epsilon+p}={1\over 4\pi}{s\over\epsilon+p}={1\over 4\pi}{(\partial p/\partial T)_\mu\over T(\partial p/\partial T)_\mu+\mu(\partial p/\partial\mu)_T}={1\over 16\pi }{ (\partial p/\partial T)_\mu\over p}\,,
\ee
where the last equality is from conformal nature $p=T^4\bar p(T/\mu)$ or $\epsilon=3p$.

From here on, we follow the convention and notation in Ref.\cite{Landsteiner:2013aba} in our subsequent analysis.
Using the scale invariance of the theory, we choose the black hole horizon radius $r_H$ to be unity, and measure all other quantities in dimensionless units by dividing with appropriate powers of $r_H$. For example, the dimensionless quantities satisfy 
\be 
m=1+q^2\,, \quad       q^2=\frac{\mu^2}{3}\,,\quad T=\frac{2m-3q^2}{2\pi}\,.
\ee
These relations allow us to characterize the plasma by a single dimensionless number, but
we should keep in mind that the truly meaningful dimensionless parameter of the conformal plasma is $T/\mu$.
Likewise, the meaningful value of the relaxation times in conformal theory is only given by the dimensionless number in unit of inverse temperature, $\tau T$.
In the holographic description, the computation of the two-point retarded correlation functions amounts to doing variation of the on-shell action up to second order in fluctuations of the values of the gauge field and the metric at the AdS boundary $r\to\infty$.
Writing the fluctuations from the above background solution as $A_M(r,t,\vec x) = A_M^{(0)}(r) + a_M(r,t,\vec x)$ and $g_{MN}(r,t,\vec x) = g_{MN}^{(0)}(r)+h_{MN}(r,t,\vec x)$, one solves the equations of motion for the fluctuations $(a_M,h_{MN})$ with the in-falling boundary condition at the horizon. The boundary values at $r\to
\infty$ are fixed and they correspond to the sources coupled to the $U(1)$ current and the energy-momentum operators in the field theory side.  By taking variations of the on-shell action with respect to these sources, we read off the retarded correlation functions. Since we are interested in computing the chiral transport coefficients arising from the P-odd correlation functions, it turns out to be enough to focus only on the shear sector, that is, we turn on only the following fluctuation components that are arranged in the single column vector $\Phi^T = (a_x, h^x_t, h^x_y, a_z, h^z_t, h^z_y)$ that only depend on $(t,y)$ coordinates. By using coordinate re-parametrization invariance and gauge invariance, one can fix the gauge by $a_r = h_{rM} = 0$.  The resulting linearized equation of motions in the Fourier space of $e^{-i\omega t+iky}$ read as
\bear
a''_i(u) + \frac{f'(u)}{f(u)} a'_i(u) + \frac{1}{4 u f^2(u)}(\omega ^2 - f(u) k^2) a_i(u) - \frac{\mu}{f(u)} h'^i_t(u) + \frac{4 i k \kappa \epsilon_{ij} \mu a_j(u)}{f(u)} &=& 0\,,\nonumber\\ 
h''^i_t(u) - \frac{h'^i_t(u)}{u} - \frac{1}{4 u f(u)}(k^2 h^i_t(u) + \omega k h^i_y(u)) - u \mu a'_i(u) &=& 0\,,\nonumber\\
h''^i_y(u) + (\frac{f'(u)}{f(u)}-\frac{1}{u})h'^i_y(u) + \frac{1}{4 u f^2(u)}(\omega ^2 h^i_y(u) + \omega k h^i_t(u)) &=& 0\,, \nonumber\\\label{EOM}
\eear
with the constraint equations
\be 
\omega h'^i_t(u) + k f(u) h'^i_y(u) - u \omega \mu a_i(u) = 0\,,\label{constrat}
\ee
where we use the new radial coordinate $u=\frac{1}{r^2}$, and $(i,j)$ runs over $x$ and $z$. 
Arranging the boundary sources at $u\to 0$ ($r\to \infty$) into a column vector \be\phi^{(0)} = \{a_x^{(0)}, h^{(0)x}_t, h^{(0)x}_y, a_z^{0}, h^{(0)z}_t, h^{(0)z}_y\}^T\,,\ee the solution of the equation of motion can be written in a matrix form
\be
\Phi^I(u) = F^I_J(k,u)\phi^{(0)J}\equiv F(k,u)\cdot \phi^{(0)}\,,
\ee
where $(I,J)$ runs over the six components of the column vectors we introduce in the above.

Following Ref.\cite{Landsteiner:2013aba}, we can compute
the $F(k,u)$ matrix using any set of six linearly independent solutions, $\{\Phi^i(k,u)\}$ ($i=1,\ldots 6$). Note that each $\Phi^i(k,u)$ is a column vector. A general solution of $\Phi(u)$ is a linear combination of  $\{\Phi^i(k,u)\}$:
\be 
\Phi(u)=\sum_{i=1}^6 \, c_i \Phi^i(k,u)\equiv [\Phi^i(k,u)]\cdot c\,,
\ee
where $c$ is the column vector of $c_i$'s, and $[\Phi^i(k,u)]$ in the above expression is a matrix whose $(I,i)$ element is the $I$'th element of the column vector $\Phi^i(k,u)$.
The constant $c_i$'s are determined by matching the boundary condition at $u\to 0$ as
\be
\phi^{(0)}=[\Phi^i(k,0)]\cdot c\,,
\ee
which can be solved by 
\be
c= ([\Phi^i(k,0)])^{-1}\cdot \phi^{(0)}\,,
\ee
and we have
\be
\Phi(u)= [\Phi^i(k,u)]\cdot c=[\Phi^i(k,u)]\cdot[\Phi^i(k,0)]^{-1} \cdot \phi^{(0)}\,,
\ee
which gives the $F$ matrix as
\be
F(k,u)=[\Phi^i(k,u)]\cdot[\Phi^i(k,0)]^{-1}\,.
\ee
In terms of the above solution of the equation of motion, the holographic renormalized action up to second order in fluctuations is given by
\be 
\delta S^{(2)} = \int \frac{d^4 k}{(2\pi)^4}\left( \Phi^T (-k)\cdot A\cdot \Phi'(k) + \Phi^T(-k)\cdot B \Phi(k) \right)\bigg|_{u\to 0}\,,
\ee
where the derivative is along the radial direction, and the matrix $A$ is  
\be 
A = \frac{1}{16 \pi G}{\rm Diag}\left( f(u), -\frac{1}{u}, \frac{f(u)}{u}, f(u), -\frac{1}{u}, \frac{f(u)}{u}\right)\,.
\ee
The matrix $B$ is the counter term from the holographic renormalization, but its components vanish for our P-odd correlation functions. From these,
one can read off the retarded correlation functions as 
\be 
G_R = -2 \lim_{u\to 0} \left(A\cdot(F(k,u))'+B \right)\,.
\ee
\begin{figure}[t]
	\centering
	\includegraphics[width=7cm]{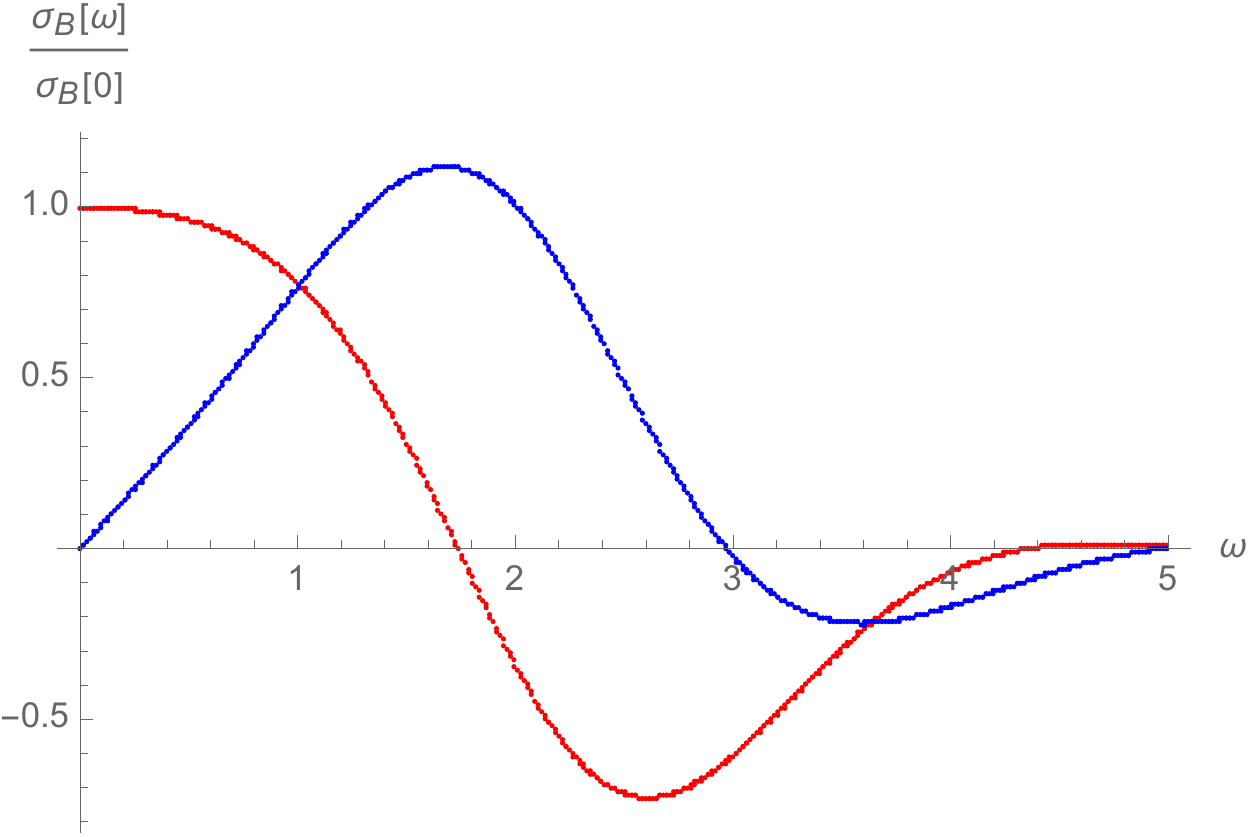} 
 \caption{The real (red) and imaginary (blue) parts of $\tilde\sigma_B(k)$ with $2\pi T/\mu=95$ when $|\bm{k}|=0.1$ as a function of frequency. \label{fig1}}
\end{figure}

With all these ingredients at hand, the numerical computation of the retarded correlation functions can be done. In the zero frequency limit, the leading order chiral transport coefficients are reproduced 
\be 
 \sigma_B = \frac{\kappa}{2 \pi G}\mu\,,\quad  \sigma_V = \sigma_\epsilon^B={1\over 2}\frac{\kappa}{2 \pi G} \mu^2\,,\quad \sigma^V_\epsilon = \frac{2}{3} \frac{\kappa}{2 \pi G}\mu^3\,.
\ee
It is more difficult to solve the coupled second order differential equations of motion at non-zero frequency.
The system of our differential equations presents a regular singularity at $u=1$, and the Frobenius method tells us that we can isolate the singular part and the rest part has a regular series expansion around $u=1$. Out of two possible singular parts, we select the one corresponding to the in-falling boundary condition as
\bear
a^i(u) &=& (1-u)^{- i \omega /(4 \pi T)}b^i(u)\,,\nonumber\\
h^i_t(u) &=& (1-u)^{- i \omega /(4 \pi T)+1}H^i_t(u)\,,\nonumber\\ 
h^i_y(u) &=& (1-u)^{- i \omega /(4 \pi T)+1}H^i_y(u)\,,
\eear
where $(b^i,H^i_{t,y})$ have regular power series expansion around $u=1$,
\bear
b^i(u) &=& \sum_{n=0} b^i_n(1-u)^n\,,\nonumber\\
H^i_{t}(u) &=& \sum_{n=0}H^i_{t n}(1-u)^n\,,\nonumber\\
H^i_y(u)&=& \sum_{n=0}H^i_{y n}(1-u)^n\,.
\eear
\begin{figure}[t]
	\centering
	\includegraphics[width=7cm]{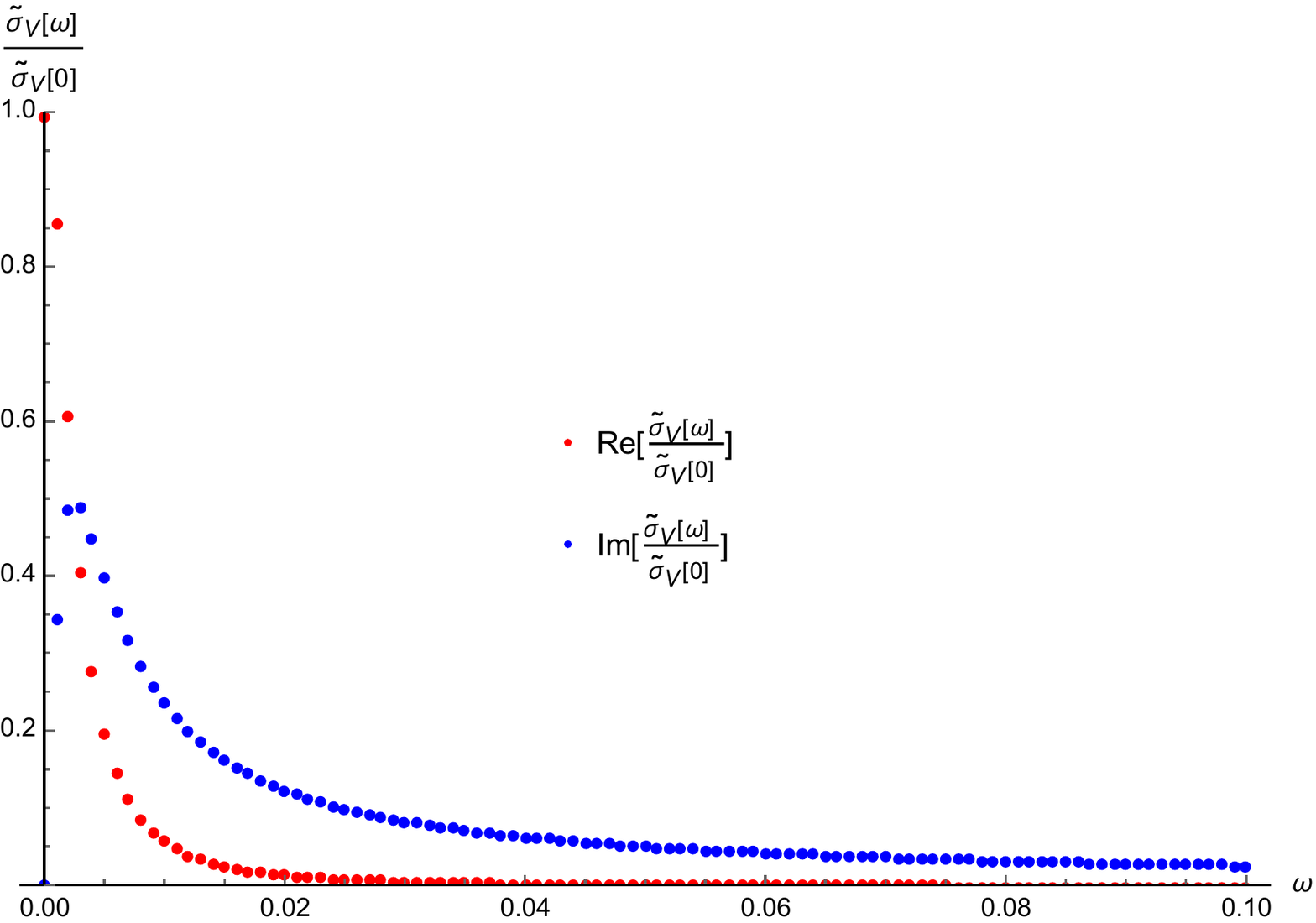} 
	\caption{The real (red) and imaginary (blue) parts of $\tilde\sigma_V(k)$ with $2\pi T/\mu=95$ when $|\bm{k}|=0.1$ as a function of frequency.\label{fig1b}}
\end{figure}
The equations of motion determine all these series coefficients in terms of the boundary values at $u=1$, that is, the six initial coefficients $\{b^i_0, H^i_{t 0}, H^i_{y 0} \}$. For example, $\{b^i(u=1)'=-b^i_1, H^i_t(u=1)'=-H^i_{t 1}, H^i_y(u=1)'=-H^i_{y 1} \}$ are given in terms of these six initial data. This gives us six linearly independent solutions $\Phi^i(k,u)$ that we need. Using these information, one can start the numerical solution of the second order differential equation slightly away from the horizon singularity $u=1+\epsilon$ with $\epsilon\ll 1$, which ensures a numerical stability.
One subtlety is that we have six second order differential equations (\ref{EOM}) and two constraints (\ref{constrat}), which tells us that the six functions $(b^i,H^i_{t,y})$ are not independent of each other. The two constraints fix the metric fluctuation $H^i_{y0}$ at the horizon in terms of  $b^i_0$ and $H^i_{t0}$ as 
\be 
H^i_{y0} = \frac{3(4 \pi T i + \omega)}{k(\mu^2-6)} H^i_{t0} + \frac{12 i \pi T \mu}{k (\mu^2 -6)}b^i_0\,.
\ee
Due to this constraint, we can find only four linearly independent solutions. The remaining two can be trivially constructed from the gauge transformations as done in Ref.\cite{Landsteiner:2013aba}. 
\begin{figure}[t]
     \centering 
     \includegraphics[width=5cm]{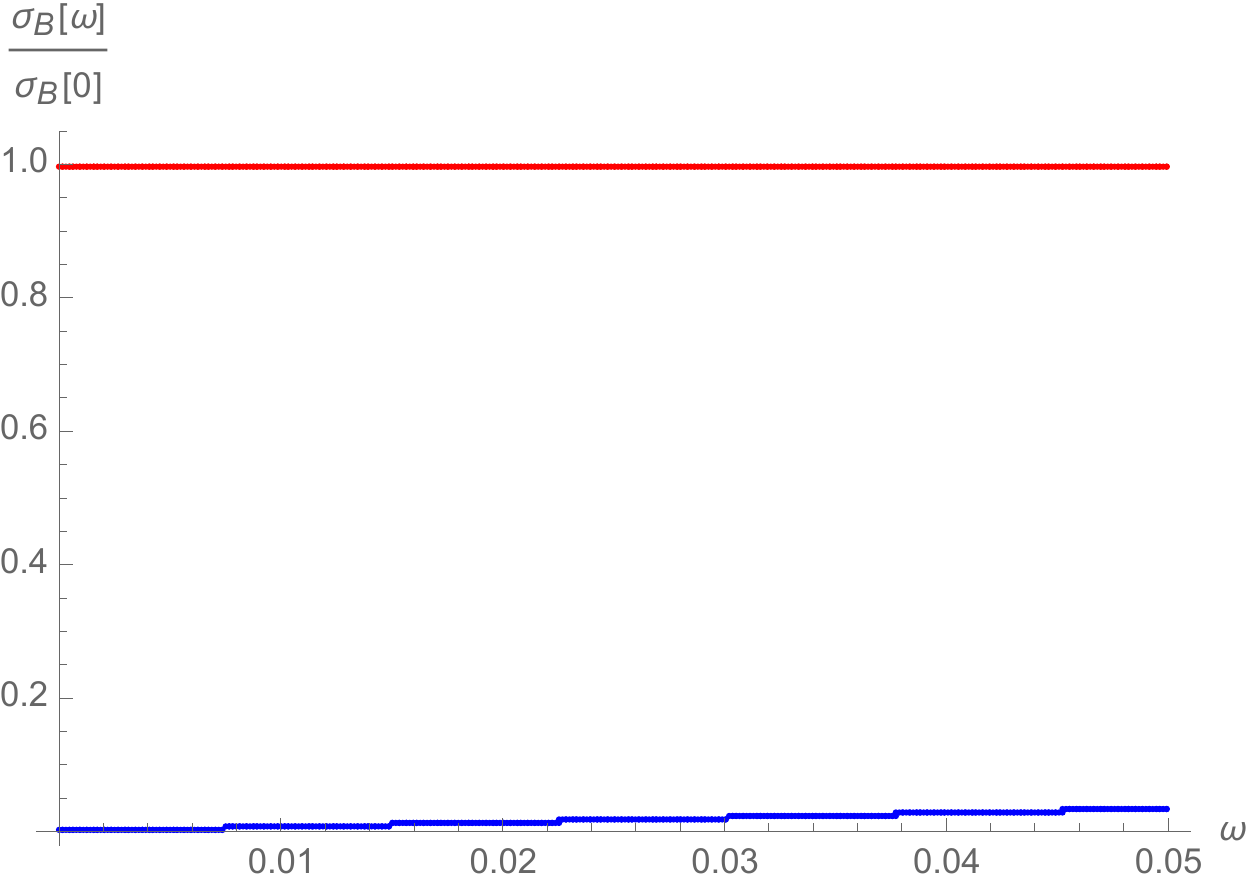} \quad \includegraphics[width=5cm]{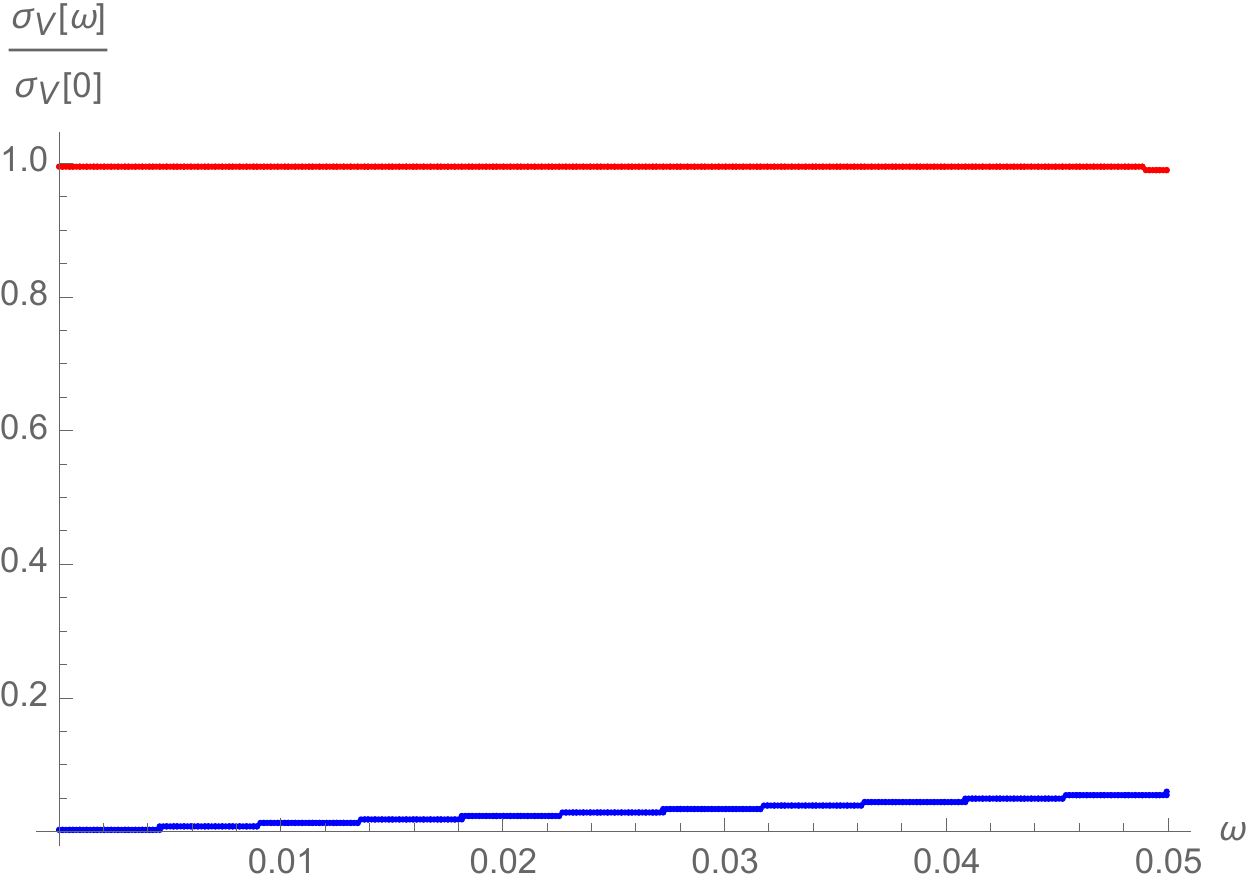}\\\includegraphics[width=5cm]{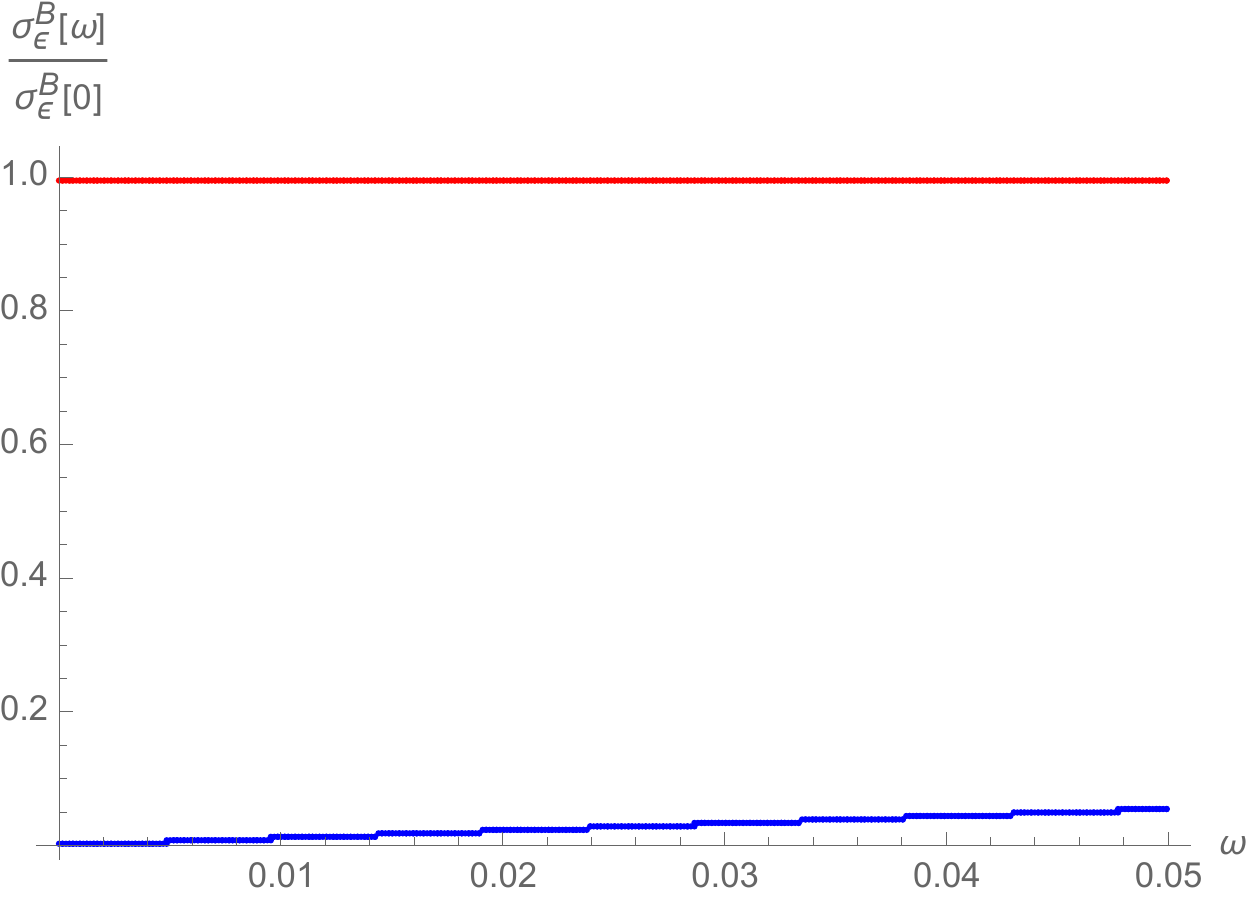} \quad \includegraphics[width=5cm]{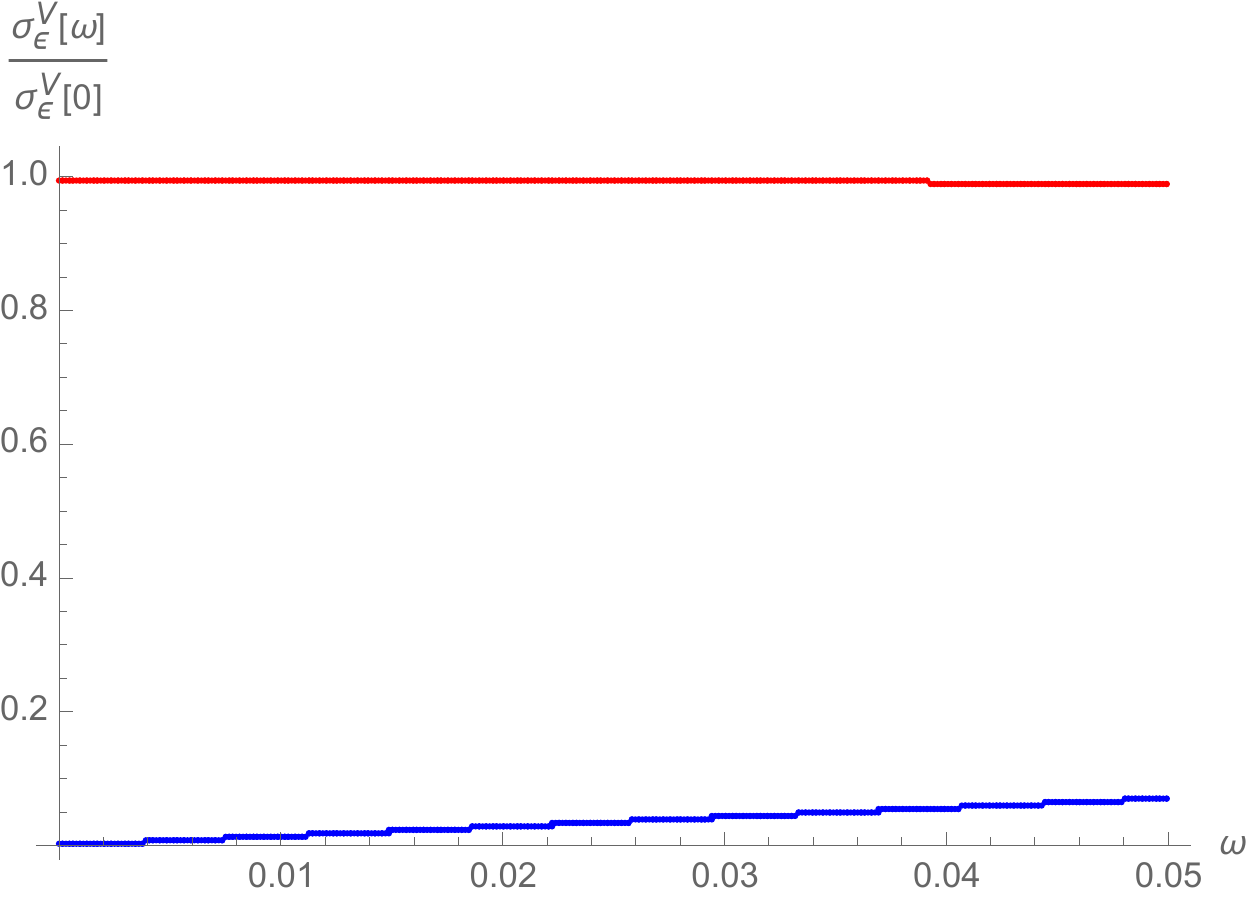} 
     \caption{The frequency-dependent chiral transport coefficients that are obtained by appropriately removing the shear diffusion pole structures from the retarded correlation functions. Red and blue curves the real and the imaginary parts respectively. \label{fig2}}
\end{figure}


In Figure \ref{fig1} we show the numerical evaluation of the retarded correlation function $\tilde\sigma_B(k)$ with $|\bm {k}|=0.1$ as a function of frequency $\omega$ with the value of $2\pi T/\mu=95$ corresponding to experimentally relevant values of $T=160$ MeV and $\mu=10$ MeV. A similar result has been previously computed in Refs.\cite{Yee:2009vw,Landsteiner:2013aba}.
In Figure \ref{fig1b} we plot $\tilde\sigma_V(k)$ with the same parameters. As observed previously in Ref.\cite{Landsteiner:2013aba}, this $\tilde\sigma_V(k)$ gets highly suppressed for small $k$ due to the shear diffusion pole structure.

However, once we strip off the diffusion pole structure as described in the section \ref{Kubo}, the resulting quantity 
\be
\sigma_V(\omega)\equiv  \lim_{\bm k\to 0}\left(\frac{\omega + i \gamma_\eta \bm{k}^2}{i\gamma_\eta \bm{k}^2}\tilde\sigma_V(k)+\frac{n}{\epsilon + p} \sigma^V_\epsilon (1 + i \omega \tau^V_\epsilon) \frac{\omega}{\omega + i \gamma_\eta \bm{k}^2}\right)\,,
\ee
has the expected dependence in small, but a finite $\omega$: especially, the imaginary part is linear in $\omega$ with a nice finite slope that is identified as the relaxation time $\tau_V$. This, together with the similar quantities $\sigma_\epsilon^{B/V}(\omega)$ that are obtained after stripping off the diffusion pole structures, are shown in Figure \ref{fig2}.
These quantities can safely be called the frequency-dependent chiral transport coefficients that appear in the {\it constitutive relations} of the chiral transport phenomena.

\begin{figure}[t]
	\centering
    \includegraphics[width=10cm]{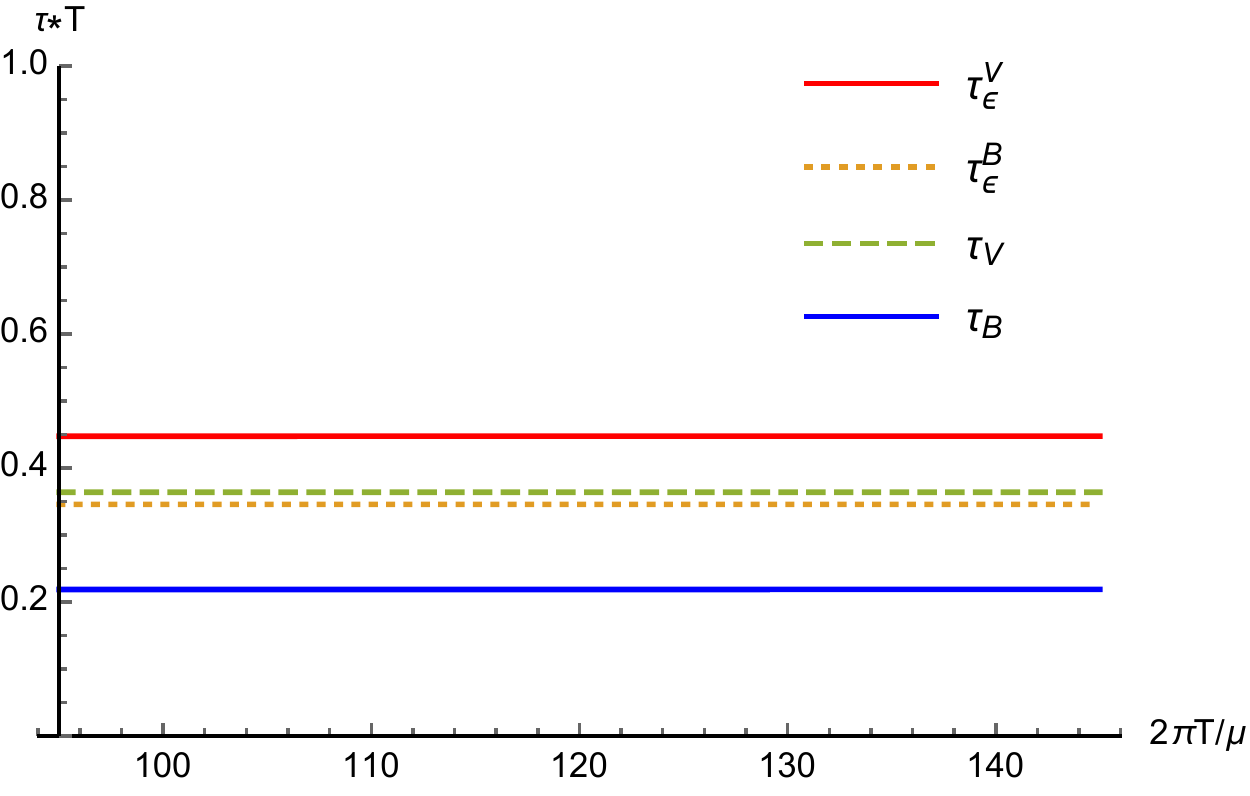}
	\caption{The numerical result for the relaxation times of chiral transport phenomena as a function of $\tau = \frac{2\pi T}{\mu}$.  \label{fig3}}
\end{figure}

In Figure \ref{fig3}, we plot the extracted relaxation times of the chiral transport phenomena, $\tau_{B/V}$ and $\tau_\epsilon^{B/V}$, as a function of the dimensionless parameter $2\pi T/\mu$.
Up to our numerical uncertainty, the result confirms our expectation of $\tau_V=\tau_\epsilon^B$ dictated by T-invariance.


\section{ Conclusion \label{sec4}}

In this work, we compute the dynamical time scale of chiral transport phenomena, that characterizes how fast an off-equilibrium condition relaxes to the equilibrium configuration that is dictated by chiral anomaly, in the strongly coupled regime using the AdS/CFT correspondence. The microscopic dynamics responsible for these relaxation to equilibrium is closely related to the dynamics of spin alignment of quasi-particles in weakly coupled regime. Although we compute these relaxation times in the strongly coupled regime using the AdS/CFT correspondence, our result should be a useful proxy for this important time scale in the real QCD plasma.

More practically, our relaxation times can be used in a Israel-Stewart type treatment of the chiral transport phenomena in realistic hydrodynamics simulations of time-dependent background of the relativistic heavy-ion collisions. We clarify some issues related to the time-dependent Chiral Vortical Effect, by identifying correct Kubo relations for the relaxation times of chiral transport coefficients. With our Kubo relations, the relaxation times for all chiral transport phenomena, including the Chiral Vortical Effect, are finite and well-defined. This should be the case, since the underlying microscopic dynamics of Chiral Vortical Effect is the dynamics of spin-polarization that must have a finite dynamical time scale to achieve the equilibrium configuration.

We also find interesting consequences of imposing time-reversal invariance on the P-odd thermal noise fluctuation correlation functions, that are related to chiral transport coefficients via fluctuation-dissipation relation. Especially we find $\tau_V=\tau_\epsilon^B$ which is confirmed in our numerical computation in the AdS/CFT correspondence.
Our discussion also sheds a light on the origin of the previously observed equality $\sigma_V=\sigma_\epsilon^B$ in terms of time-reversal invariance.

The relaxation times that we consider in this work are examples of second order chiral transport coefficients.
It would be interesting to compute these and the other second order chiral transport coefficients in weakly coupled regime in perturbative QCD, as was done in Ref.\cite{Jimenez-Alba:2015bia} for $\tau_B$.

\vskip 1cm \centerline{\large \bf Acknowledgment} \vskip 0.5cm

We thank Dima Kharzeev, Karl Landsteiner and Misha Stephanov for helpful discussions.
This work is supported by the U.S. Department of Energy, Office of Science, Office of Nuclear Physics, with the grant No. DE-SC0018209 and within the framework of the Beam Energy Scan Theory (BEST) Topical Collaboration.

 \vfil

\end{document}